\documentstyle[prd,aps,preprint,tighten,epsfig]{revtex}

\begin{document}

\draft

\title{Friedberg-Lee Symmetry Breaking and Its
Prediction for $\theta^{}_{13}$}
\author{{\bf Shu Luo} \thanks{E-mail: luoshu@mail.ihep.ac.cn} ~ and
~ {\bf Zhi-zhong Xing} \thanks{E-mail: xingzz@mail.ihep.ac.cn} }
\address{
CCAST (World Laboratory), P.O. Box 8730, Beijing 100080,
China \\
and Institute of High Energy Physics, Chinese Academy of Sciences,
\\
P.O. Box 918, Beijing 100049, China}

\maketitle

\begin{abstract}
We consider an effective Majorana neutrino mass operator with the
Friedberg-Lee symmetry; i.e., it is invariant under the
transformation $\nu^{}_\alpha \rightarrow \nu^{}_\alpha + z$ (for
$\alpha = e, \mu, \tau$) with $z$ being a space-time independent
constant element of the Grassmann algebra. We show that this new
flavor symmetry can be broken in such a nontrivial way that the
lightest neutrino remains massless but an experimentally-favored
neutrino mixing pattern is achievable. In particular, we get a
novel prediction for the unknown neutrino mixing angle
$\theta^{}_{13}$ in terms of two known angles: $\sin\theta^{}_{13}
= \tan\theta^{}_{12} |(1- \tan\theta^{}_{23})/ (1+
\tan\theta^{}_{23})|$. The model can simply be generalized to
accommodate CP violation and be combined with the seesaw
mechanism.
\end{abstract}

\pacs{PACS number(s): 14.60.Lm, 14.60.Pq, 95.85.Ry}


Recent solar \cite{SNO}, atmospheric \cite{SK}, reactor \cite{KM}
and accelerator \cite{K2K} neutrino experiments have convincingly
verified the hypothesis of neutrino oscillations. The latter can
naturally happen if neutrinos are slightly massive and lepton
flavors are not conserved. The mixing of three lepton families is
described by a $3\times 3$ unitary matrix $V$, whose nine elements
are usually parameterized in terms of three rotation angles
($\theta^{}_{12}$, $\theta^{}_{23}$, $\theta^{}_{13}$) and three
CP-violating phases ($\delta$, $\rho$, $\sigma$) \cite{FX01}:
\begin{equation}
V = \left( \matrix{ c^{}_{12}c^{}_{13} & s^{}_{12}c ^{}_{13} &
s^{}_{13} e^{-i\delta} \cr -s^{}_{12}c^{}_{23}
-c^{}_{12}s^{}_{23}s^{}_{13} e^{i\delta} & c^{}_{12}c^{}_{23}
-s^{}_{12}s^{}_{23}s^{}_{13} e^{i\delta} & s^{}_{23}c^{}_{13} \cr
s^{}_{12}s^{}_{23} -c^{}_{12}c^{}_{23}s^{}_{13} e^{i\delta} &
-c^{}_{12}s^{}_{23} -s^{}_{12}c^{}_{23}s^{}_{13} e^{i\delta} &
c^{}_{23}c^{}_{13} } \right) \left ( \matrix{e^{i\rho } & 0 & 0
\cr 0 & e^{i\sigma} & 0 \cr 0 & 0 & 1 \cr} \right ) \;
\end{equation}
with $c^{}_{ij} \equiv \cos\theta_{ij}$ and $s^{}_{ij} \equiv
\sin\theta_{ij}$ (for $ij=12,23$ and $13$). A global analysis of
current neutrino oscillation data yields $30^\circ \leq
\theta_{12} \leq 38^\circ$, $36^\circ \leq \theta_{23} \leq
54^\circ$ and $0^\circ \leq \theta_{13} < 10^\circ$ at the $99\%$
confidence level \cite{Vissani}, but three phases of $V$ remain
entirely unconstrained. While the absolute mass scale of three
neutrinos is not yet fixed, their two mass-squared differences
have already been determined to a quite good degree of accuracy
\cite{Vissani}: $\Delta m^2_{21} \equiv m^2_2 - m^2_1 = (7.2
\cdot\cdot\cdot 8.9) \times 10^{-5} ~{\rm eV}^2$ and $\Delta
m^2_{32} \equiv m^2_3 - m^2_2 = \pm (2.1 \cdot\cdot\cdot 3.1)
\times 10^{-3} ~{\rm eV}^2$ at the $99\%$ confidence level. The
on-going and forthcoming neutrino oscillation experiments aim to
measure the sign of $\Delta m^2_{32}$, the magnitude of
$\theta^{}_{13}$ and even the CP-violating phase $\delta$.

How to understand the smallness of $\theta^{}_{13}$ and the
largeness of $\theta^{}_{12}$ and $\theta^{}_{23}$ is a real
challenge. So far many neutrino mass models have been proposed
\cite{Review}. Some of them follow such a guiding principle: there
exists an underlying flavor symmetry in the lepton sector and its
spontaneous or explicit breaking gives rise to the observed
pattern of $V$. This is certainly a reasonable starting point for
model building, and it might even shed light on the true flavor
structures of leptons and quarks.

The purpose of this paper is just to follow the above-mentioned
guideline to explore a simple and testable correlation between the
neutrino mass spectrum and the neutrino mixing pattern. In the
basis where the charged-lepton mass matrix is diagonal, we
hypothesize that the effective Majorana neutrino mass operator is
of the form
\begin{eqnarray}
{\cal L}^{}_{\rm mass} & = & \frac{1}{2} \left [ a
(\overline{\nu^{}_{\tau \rm L}}- \overline{\nu^{}_{\mu \rm L}})
(\nu^{\rm c}_{\tau \rm L} - \nu^{\rm c}_{\mu \rm L}) + b
(\overline{\nu^{}_{\mu \rm L}} - \overline{\nu^{}_{e \rm L}})
(\nu^{\rm c}_{\mu \rm L} - \nu^{\rm c}_{e \rm L}) \right .
\nonumber \\
&& \left. ~~~~~~~~~~~~~~~~~~~~~~~~~~~~~~~~~~ + c
(\overline{\nu^{}_{e \rm L}} - \overline{\nu^{}_{\tau \rm L}})
(\nu^{\rm c}_{e \rm L} - \nu^{\rm c}_{\tau \rm L}) \right ] ~ + ~
{\rm h.c.} \; ,
\end{eqnarray}
where $a$, $b$ and $c$ are in general complex, and $\nu^{\rm
c}_{\alpha \rm L} \equiv C\overline{\nu^{}_{\alpha \rm L}}^T$ (for
$\alpha =e$, $\mu$, $\tau$). A salient feature of ${\cal
L}^{}_{\rm mass}$ is its translational symmetry; i.e., ${\cal
L}^{}_{\rm mass}$ is invariant under the transformation
$\nu^{}_\alpha \rightarrow \nu^{}_\alpha + z$ with $z$ being a
space-time independent constant element of the Grassmann algebra.
Note that this new kind of flavor symmetry was first introduced by
Friedberg and Lee in Ref. \cite{Lee} to describe the Dirac
neutrino mass operator. Here we apply the same symmetry to the
case of Majorana neutrinos and then reveal a completely new way to
break it. Corresponding to Eq. (2), the Majorana neutrino mass
matrix $M^{}_\nu$ reads
\begin{equation}
M^{}_\nu = \left ( \matrix{ b+c & -b & -c \cr -b & a+b & -a \cr -c
& -a & a+c \cr} \right ) \; .
\end{equation}
The diagonalization of $M^{}_\nu$ is straightforward: $V^\dagger
M^{}_\nu V^* = \overline{M}^{}_\nu$, where $V$ is just the
neutrino mixing matrix, and $\overline{M}^{}_\nu = {\rm Diag} \{
m^{}_1, m^{}_2, m^{}_3 \}$ with $m^{}_i$ (for $i=1,2,3$) being the
neutrino masses. From Eq. (3) together with the parametrization of
$V$ in Eq. (1), it is easy to verify
\begin{equation}
{\rm Det}(M^{}_\nu) = {\rm Det} \left ( V \overline{M}^{}_\nu V^T
\right ) = {\rm Det} (\overline{M}^{}_\nu) \left [ {\rm Det}(V)
\right ]^2 = m^{}_1 m^{}_2 m^{}_3 e^{2 i \left (\rho + \sigma
\right )} = 0 \; .
\end{equation}
This result, which is an immediate consequence of the
Friedberg-Lee (FL) symmetry in ${\cal L}^{}_{\rm mass}$, implies
that one of the three neutrinos must be massless. In the FL model
\cite{Lee} such a flavor symmetry is broken by an extra term of
the form $m^{}_0 (\overline{\nu}^{}_e \nu^{}_e +
\overline{\nu}^{}_\mu \nu^{}_\mu + \overline{\nu}^{}_\tau
\nu^{}_\tau)$ added into the Dirac neutrino mass operator, hence
all the three neutrinos are massive.

One may wonder whether it is possible to break the FL symmetry in
${\cal L}^{}_{\rm mass}$ but keep $m^{}_1 =0$ or $m^{}_3 =0$
unchanged?
\footnote{Because of $m^{}_2 > m^{}_1$ obtained from the solar
neutrino oscillation data \cite{SNO}, it makes no sense to
consider the $m^{}_2 =0$ case.}
We find that the simplest way to make this possibility realizable
is to transform one of the neutrino fields $\nu^{}_\alpha$ into
$\kappa^* \nu^{}_\alpha$ with $\kappa \neq 1$. Taking $\nu^{}_e
\rightarrow \kappa^* \nu^{}_e$ for example, the resultant Majorana
neutrino mass operator takes the form
\begin{eqnarray}
{\cal L}^\prime_{\rm mass} & = & \frac{1}{2} \left [ a
(\overline{\nu^{}_{\tau \rm L}}- \overline{\nu^{}_{\mu \rm L}})
(\nu^{\rm c}_{\tau \rm L} - \nu^{\rm c}_{\mu \rm L}) + b
(\overline{\nu^{}_{\mu \rm L}} - \kappa \overline{\nu^{}_{e \rm
L}}) (\nu^{\rm c}_{\mu \rm L} - \kappa \nu^{\rm c}_{e \rm L})
\right . \nonumber \\
&& \left . ~~~~~~~~~~~~~~~~~~~~~~~~~~~~~~~~~~ + c ( \kappa
\overline{\nu^{}_{e \rm L}} - \overline{\nu^{}_{\tau \rm L}}) (
\kappa \nu^{\rm c}_{e \rm L} - \nu^{\rm c}_{\tau \rm L}) \right ]
~ + ~ {\rm h.c.} \; .
\end{eqnarray}
Accordingly, the neutrino mass matrix is given by
\begin{equation}
M^\prime_\nu = \left ( \matrix{ \kappa^2 (b+c) & - \kappa b & -
\kappa c \cr - \kappa b & a+b & -a \cr - \kappa c & -a & a+c \cr}
\right ) \; .
\end{equation}
We are then left with ${\rm Det}(M^\prime_\nu) = \kappa^2 {\rm
Det}(M^{}_\nu) =0$, which is independent of the magnitude and
phase of $\kappa$. Thus we obtain either $m^{}_1 =0$ or $m^{}_3
=0$ from $M^\prime_\nu$. Our next step is to show that a generic
bi-large neutrino mixing pattern, which is compatible very well
with current experimental data, can be derived from
$M^\prime_\nu$.

Let us focus on the $m^{}_1 =0$ case
\footnote{We find that the $m^{}_3 =0$ case is actually
disfavored, if we intend to achieve $\theta^{}_{23} = \pi/4$ and
$\theta^{}_{13} =0$ from $M^\prime_\nu$ in the leading-order
approximation.},
in which $M^\prime_\nu$ is diagonalized by the transformation
$V^\dagger M^\prime_\nu V^* = \overline{M}^{}_\nu$ with
$\overline{M}^{}_\nu = {\rm Diag}\{0, m^{}_2, m^{}_3 \}$. As the
best-fit values of the atmospheric and CHOOZ neutrino mixing
angles are $\theta^{}_{23} =\pi/4$ and $\theta^{}_{13} =0$
respectively \cite{Vissani}, we may decompose the neutrino mixing
matrix $V$ into a product of three special unitary matrices: $V =
U RP$, where
\begin{eqnarray}
U & = & \left ( \matrix{ &  & 0 \cr {\bf u}^{}_1 & {\bf u}^{}_2 &
\frac{1}{\sqrt{2}} \cr & & \frac{-1}{\sqrt{2}} \cr} \right ) \; ,
\nonumber \\
R & = & \left ( \matrix{1 & ~~ 0 ~~ & 0 \cr 0 & \hat{c} &
\tilde{s} \cr 0 & - \tilde{s}^* & \hat{c}^* \cr} \right ) \; ,
\end{eqnarray}
and $P = {\bf 1} e^{i\gamma}$ with the definitions $\hat{c} \equiv
\cos\theta ~e^{i\phi}$ and $\tilde{s} \equiv \sin\theta ~
e^{i\varphi}$. Note that ${\bf u}^*_1$ is a column vector
associated with $m^{}_1 =0$ (i.e., $M^\prime_\nu {\bf u}^*_1 = 0$
holds). This observation, together with the unitarity of $U$,
allows us to obtain
\begin{eqnarray}
{\bf u}^{}_1 & = & \frac{1}{\sqrt{2|\kappa|^2 +1}} \left (
\matrix{ ~ 1 ~\cr \kappa^* \cr \kappa^* \cr} \right ) \; ,
\nonumber \\
{\bf u}^{}_2 & = & \frac{\sqrt{2}}{\sqrt{2|\kappa|^2 +1}} \left (
\matrix{ \kappa \cr \frac{-1}{2} \cr \frac{-1}{2} \cr} \right ) \;
.
\end{eqnarray}
One can see that $U$ is only dependent on $\kappa$, a free
parameter characterizing the strength of FL symmetry breaking in
${\cal L}^\prime_{\rm mass}$. Apparently, $V^\dagger M^\prime_\nu
V^* = P^\dagger R^\dagger (U^\dagger M^\prime_\nu U^*) R^* P^*$
holds, where
\begin{equation}
U^\dagger M^\prime_\nu U^* = \frac{1}{2} \left ( \matrix{ 0 & 0 &
0 \cr 0 & \left (b+c \right ) \left ( 2 |\kappa|^2 + 1 \right ) &
\left (c-b \right ) \sqrt{2 |\kappa|^2 + 1} \cr 0 & \left ( c-b
\right ) \sqrt{ 2 |\kappa|^2 +1} & 4a + b+c \cr} \right ) \; .
\end{equation}
If $a$, $b$ and $c$ are all complex, the phases of $U^\dagger
M^\prime_\nu U^*$ will finally be absorbed into the phases of $P$
($\gamma$) and $R$ ($\phi$ and $\varphi$).

For simplicity, we assume that $a$, $b$ and $c$ are all real. In
this case, $U^\dagger M^\prime_\nu U^*$ is a real symmetric
matrix. Hence the phases of $R$ and $P$ (i.e., $\gamma$, $\phi$
and $\varphi$) can be switched off and the rotation angle of $R$
is determined by
\begin{equation}
\tan 2\theta = \frac{\left (b-c \right ) \sqrt{ 2|\kappa|^2 +
1}}{\left (b+c \right ) |\kappa|^2 - 2 a} \; .
\end{equation}
One may observe that $b =c$, which is a clear reflection of the
$\mu$-$\tau$ permutation symmetry in ${\cal L}^\prime_{\rm mass}$
or $M^\prime_\nu$ \cite{XZZ}, simply leads to $\theta =0$ or
equivalently $\theta^{}_{23} = \pi/4$ and $\theta^{}_{13} =0$. In
addition, two non-vanishing neutrino masses are obtained from Eq.
(9) as follows:
\begin{eqnarray}
m^{}_2 & = & a + \frac{1}{2} \left (b + c\right ) \left
(|\kappa|^2 + 1 \right ) - \frac{1}{2} \sqrt{\left [ 2a - \left (b
+ c \right ) |\kappa|^2 \right ]^2 + \left (b -c \right )^2 \left
(2|\kappa|^2 + 1\right )}
\;\; , \nonumber \\
m^{}_3 & = & a + \frac{1}{2} \left (b + c\right ) \left
(|\kappa|^2 + 1 \right ) + \frac{1}{2} \sqrt{\left [ 2a - \left (b
+ c \right ) |\kappa|^2 \right ]^2 + \left (b -c \right )^2 \left
(2|\kappa|^2 + 1\right )} \;\; .
\end{eqnarray}
Nine elements of the neutrino mixing matrix $V = URP$ can be given
in terms of $\kappa$ and $\theta$:
\begin{equation}
V = \left ( \matrix{ \frac{1}{\sqrt{2|\kappa|^2 +1}} &
\frac{\sqrt{2} \; \kappa \cos\theta}{\sqrt{2|\kappa|^2 +1}} &
\frac{\sqrt{2} \; \kappa \sin\theta}{\sqrt{2|\kappa|^2 +1}} \cr
\frac{\kappa^*}{\sqrt{2|\kappa|^2 +1}} & -\frac{1}{\sqrt{2}} \left
( \frac{\cos\theta}{\sqrt{2|\kappa|^2 +1}} + \sin\theta \right ) &
~\;\; \frac{1}{\sqrt{2}} \left ( \cos\theta -
\frac{\sin\theta}{\sqrt{2|\kappa|^2 +1}} \right ) \cr
\frac{\kappa^*}{\sqrt{2|\kappa|^2 +1}} & -\frac{1}{\sqrt{2}} \left
( \frac{\cos\theta}{\sqrt{2|\kappa|^2 +1}} - \sin\theta \right ) &
-\frac{1}{\sqrt{2}} \left ( \cos\theta
+\frac{\sin\theta}{\sqrt{2|\kappa|^2 +1}} \right ) \cr} \right )
\; .
\end{equation}
After rephasing this expression of $V$ with the transformations of
charged-lepton and neutrino fields $e \rightarrow
e^{i\phi^{}_\kappa} e$, $\mu \rightarrow -\mu$, $\nu^{}_1
\rightarrow e^{i\phi^{}_\kappa} \nu^{}_1$ and $\nu^{}_3
\rightarrow -\nu^{}_3$, where $\phi^{}_\kappa \equiv \arg
(\kappa)$, we may directly compare it with the parametrization
given in Eq. (1). Then we arrive at
\begin{eqnarray}
\tan\theta^{}_{12} & = & \sqrt{2} |\kappa| \cos\theta \; ,
\nonumber \\
\tan\theta^{}_{23} & = & \left | \frac{\sqrt{2|\kappa|^2 + 1} -
\tan\theta}{\sqrt{2|\kappa|^2 + 1} + \tan\theta} \right | \; ,
\nonumber \\
\sin\theta^{}_{13} & = & \frac{\sqrt{2} |\kappa|
|\sin\theta|}{\sqrt{2|\kappa|^2 +1}} \; ,
\end{eqnarray}
where $\theta^{}_{12}$, $\theta^{}_{23}$ and $\theta^{}_{13}$ are
required to lie in the first quadrant, $\theta$ is close to zero
due to the smallness of $\theta^{}_{13}$ but it may be either
positive (in the first quadrant) or negative (in the fourth
quadrant). Furthermore, we have $\delta =0$ (when $\theta <0$) or
$\delta =\pi$ (when $\theta >0$) together with $\sigma = \pi$,
while the CP-violating phase $\rho$ is not well-defined in the
$m^{}_1 =0$ case. Thus we conclude that there is no CP violation
in this simple neutrino mass model, although its mass operator
involves a complex parameter $\kappa$.

Now that the mixing angles $\theta^{}_{12}$ and $\theta^{}_{23}$
are already known to a reasonable degree of accuracy, we can use
them to determine the unknown mixing angle $\theta^{}_{13}$ and
the unknown magnitude of $\kappa$ from Eq. (13). Indeed,
\begin{eqnarray}
\sin\theta^{}_{13} & = & \left |\frac{1 - \tan\theta^{}_{23}}{1+
\tan\theta^{}_{23}} \right | \tan\theta^{}_{12} \; .
\end{eqnarray}
This interesting expression indicates that the deviation of
$\theta^{}_{13}$ from zero is closely correlated with the
deviation of $\theta^{}_{23}$ from $\pi/4$. It is a novel
prediction of our model, which can easily be tested in the near
future. On the other hand,
\begin{eqnarray}
|\kappa| & = & \frac{\sin\theta^{}_{12}}{\sqrt{\cos
2\theta^{}_{12} + \sin 2\theta^{}_{23}}} \; .
\end{eqnarray}
Because of $m^{}_2 = \sqrt{\Delta m^2_{21}}$ and $m^{}_3 =
\sqrt{\Delta m^2_{21} + |\Delta m^2_{32}|}$ in the $m^{}_1 =0$
case, we get $m^{}_2 \approx (8.48 \cdots 9.43) \times 10^{-3}$ eV
and $m^{}_3 \approx (4.58 \cdots 5.57) \times 10^{-2}$ eV from the
$99\%$ confidence-level ranges of $\Delta m^2_{21}$ and $|\Delta
m^2_{32}|$ \cite{Vissani}. These results, together with the
experimental values of $\theta^{}_{12}$ and $\theta^{}_{23}$,
allow us to numerically constrain the model parameters via Eqs.
(10), (11) and (13). We obtain $0.019 ~{\rm eV} \lesssim a
\lesssim 0.026 ~{\rm eV}$, $0.41 \lesssim |\kappa| \lesssim 0.56$,
$|\theta| < 11.4^\circ$ and the ranges of $b$ and $c$ shown in
FIG. 1. Note that the region of $|\kappa|$ can also be achieved
from Eq. (15). In particular, $|\kappa| =1/2$ is favorable and it
implies that $U$ takes the so-called tri-bimaximal mixing pattern
\cite{TB}. The numerical dependence of $\theta^{}_{13}$ on
$\theta^{}_{12}$ and $\theta^{}_{23}$ is illustrated in FIG. 2,
from which an upper bound $\theta^{}_{13} \leq 7.1^\circ$ can be
extracted. Such a constraint on $\theta^{}_{13}$ is certainly more
stringent than $\theta^{}_{13} < 10^\circ$ obtained from a global
analysis of current neutrino oscillation data \cite{Vissani}. It
will be interesting to see whether our prediction for the
correlation between the unknown mixing angle $\theta^{}_{13}$ and
two known angles can survive the future measurements.

In the above discussions we have taken $\nu^{}_e \rightarrow
\kappa^* \nu^{}_e$ to break the FL symmetry and achieve a
realistic pattern of the neutrino mass matrix. One may similarly
consider $\nu^{}_\mu \rightarrow \kappa^* \nu^{}_\mu$ or
$\nu^{}_\tau \rightarrow \kappa^* \nu^{}_\tau$ with $\kappa \neq
1$. In either possibility it is easy to show that $m^{}_1 =0$ or
$m^{}_3 =0$ holds, but the neutrino mixing pattern turns out to be
disfavored by current experimental data. We find that there is no
way to simultaneously obtain large $\theta^{}_{23}$ and tiny
$\theta^{}_{13}$ in the $m^{}_3 = 0$ case, no matter whether
$\nu^{}_\mu \rightarrow \kappa^* \nu^{}_\mu$ or $\nu^{}_\tau
\rightarrow \kappa^* \nu^{}_\tau$ is taken. As for the $m^{}_1 =0$
case, it is straightforward to get the neutrino mixing matrix from
Eq. (12) with the interchange of its first and second rows (when
$\nu^{}_\mu \rightarrow \kappa^* \nu^{}_\mu$ is concerned) or its
first and third rows (when $\nu^{}_\tau \rightarrow \kappa^*
\nu^{}_\tau$ is concerned). We observe that $|\kappa| \sim 1$ is
required to assure $\theta^{}_{23} \sim \pi/4$ and $\theta^{}_{13}
\sim 0$ in the leading-order approximation, either for $\nu^{}_\mu
\rightarrow \kappa^* \nu^{}_\mu$ or for $\nu^{}_\tau \rightarrow
\kappa^* \nu^{}_\tau$. But $|\kappa| \sim 1$ will give rise to an
excessively large value of $\theta^{}_{12}$ (e.g., $\theta^{}_{12}
> \pi/4$), which has been ruled out by the solar neutrino
oscillation data. Hence neither $\nu^{}_\mu \rightarrow \kappa^*
\nu^{}_\mu$ nor $\nu^{}_\tau \rightarrow \kappa^* \nu^{}_\tau$
with $\kappa \neq 1$, which automatically breaks the $\mu$-$\tau$
permutation symmetry, is favored to reproduce the exactly or
approximately tri-bimaximal neutrino mixing pattern.

Although the discussions from Eq. (10) to Eq. (15) are based on
the assumption of real $a$, $b$ and $c$, they can easily be
extended to the case of complex $a$, $b$ and $c$ in order to
accommodate CP violation. For simplicity of illustration, here we
assume that $a$ remains real but $b = c^*$ is complex. One may
then simplify the expression of $U^\dagger M^\prime_\nu U^*$ in
Eq. (9) by taking into account $b+c = 2{\rm Re}(b)$ and $b-c = 2i
{\rm Im}(b)$. After an analogous calculation, we obtain the
neutrino mixing matrix
\begin{equation}
V = \left ( \matrix{ \frac{1}{\sqrt{2|\kappa|^2 +1}} & i
\frac{\sqrt{2} \; \kappa \cos\theta}{\sqrt{2|\kappa|^2 +1}} & i
\frac{\sqrt{2} \; \kappa \sin\theta}{\sqrt{2|\kappa|^2 +1}} \cr
\frac{\kappa^*}{\sqrt{2|\kappa|^2 +1}} & -\frac{1}{\sqrt{2}} \left (
i \frac{\cos\theta}{\sqrt{2|\kappa|^2 +1}} + \sin\theta \right ) &
~\;\; \frac{1}{\sqrt{2}} \left ( \cos\theta - i
\frac{\sin\theta}{\sqrt{2|\kappa|^2 +1}} \right ) \cr
\frac{\kappa^*}{\sqrt{2|\kappa|^2 +1}} & -\frac{1}{\sqrt{2}} \left (
i \frac{\cos\theta}{\sqrt{2|\kappa|^2 +1}} - \sin\theta \right ) &
-\frac{1}{\sqrt{2}} \left ( \cos\theta + i
\frac{\sin\theta}{\sqrt{2|\kappa|^2 +1}} \right ) \cr} \right ) \; ,
\end{equation}
where $\theta$ is given by $\tan 2\theta = - {\rm Im}(b) \sqrt{
2|\kappa|^2 + 1}/\left [a + {\rm Re}(b)\left (|\kappa|^2 + 1
\right ) \right ]$. Two immediate but important observations are
in order:
\begin{itemize}
\item       In this simple scenario $V$ contains two nontrivial
CP-violating phases: $\delta = - \pi/2$ (when $\theta <0$) or
$\delta = \pi/2$ (when $\theta
>0$) and $\sigma = - \pi/2$. Both of them are attributed to the purely
imaginary term $b-c$. The Jarlskog invariant of CP violation
\cite{J} reads ${\cal J} = |\kappa|^2 |\sin 2\theta|/[2
(2|\kappa|^2 +1)^{3/2}]$. A numerical analysis yields $0.41
\lesssim |\kappa| \lesssim 0.57$ and $|\theta| < 19.4^{\circ}$.
Thus we arrive at ${\cal J} \lesssim 0.041$. It is likely to
measure ${\cal J}\sim {\cal O}(10^{-2})$ in the future
long-baseline neutrino oscillation experiments.

\item      $\tan \theta^{}_{23} = 1$ or $\theta^{}_{23} =\pi/4$
can be achieved, although the neutrino mass operator ${\cal
L}^\prime_{\rm mass}$ does not possess the exact $\mu$-$\tau$
symmetry. The reason is simply that $|b| = |c|$ holds in our
scenario. In other words, the phase difference between $b$ and $c$
signifies a kind of {\it soft} $\mu$-$\tau$ symmetry breaking
which can keep $\theta^{}_{23} =\pi/4$ but cause $\theta^{}_{13}
\neq 0$ \cite{XZZ}. Note that Eq. (16) also yields
$\sin\theta^{}_{13} = \sqrt{2} |\kappa|
|\sin\theta|/\sqrt{2|\kappa|^2 +1}~$ and $\tan\theta^{}_{12} =
\sqrt{2} |\kappa| \cos\theta$, exactly identical to the
expressions given in Eq. (13).
\end{itemize}
It is worth mentioning that the present scenario has the same
number of free parameters as the previous one. Taking account of
current experimental data on $\Delta m^2_{21}$, $\Delta m^2_{32}$,
$\theta^{}_{12}$ and $\theta^{}_{13}$, we arrive at $0.026 ~{\rm
eV} \lesssim a \lesssim 0.032 ~{\rm eV}$, $-0.010 ~{\rm eV}
\lesssim {\rm Re}(b) \lesssim -0.005 ~{\rm eV}$ and $-0.013 ~{\rm
eV} \lesssim {\rm Im}(b) \lesssim 0.013 ~{\rm eV}$ from a
straightforward calculation.

Finally we point out that it is possible to derive the Majorana
neutrino mass operator ${\cal L}^\prime_{\rm mass}$ from the
minimal seesaw model (MSM) \cite{MSM}, a canonical extension of
the standard model with only two heavy right-handed Majorana
neutrinos. The neutrino mass term in the MSM can be written as
\begin{equation}
-{\cal L}^{}_{\rm MSM} \; = \; \frac{1}{2} \overline{\left
(\nu^{~}_{\rm L}, ~N^{\rm c}_{\rm R} \right )} \left ( \matrix{
{\bf 0} & M^{~}_{\rm D} \cr M^T_{\rm D} & M^{~}_{\rm R} \cr}
\right ) \left ( \matrix{ \nu^{\rm c}_{\rm L} \cr N^{}_{\rm R}
\cr} \right ) ~ + ~ {\rm h.c.} \; ,
\end{equation}
where $\nu^{~}_{\rm L}$ and $N^{}_{\rm R}$ denote the column
vectors of $(\nu^{~}_e, \nu^{~}_\mu, \nu^{~}_\tau)^{~}_{\rm L}$
and $(N^{~}_1, N^{~}_2)^{}_{\rm R}$ fields, respectively. Provided
the mass scale of $M^{}_{\rm R}$ is considerably higher than that
of $M^{}_{\rm D}$, one may obtain the effective (left-handed)
Majorana neutrino mass matrix $M^\prime_\nu$ from Eq. (17) via the
well-known seesaw mechanism \cite{SS}: $M^\prime_\nu = M^{}_{\rm
D} M^{-1}_{\rm R} M^T_{\rm D}$. As $M^{}_{\rm R}$ is of rank 2,
${\rm Det}(M^\prime_\nu) =0$ holds and $m^{}_1 =0$ (or $m^{}_3
=0$) is guaranteed. We find that the expression of $M^\prime_\nu$
given in Eq. (6) can be reproduced from $M^{}_{\rm D}$ and
$M^{}_{\rm R}$ if they take the following forms:
\begin{eqnarray}
M^{}_{\rm D} & = & \Lambda^{}_{\rm D} \left ( \matrix{\kappa & 0
\cr -1 & -1 \cr 0 & 1 \cr} \right ) \; , \nonumber \\
M^{}_{\rm R} & = & \frac{\Lambda^2_{\rm D}}{ab + bc + ca} \left (
\matrix{~ a+c ~ & c \cr c & ~ b+c ~ \cr} \right ) \; ,
\end{eqnarray}
where $\Lambda^{}_{\rm D}$ characterizes the mass scale of
$M^{}_{\rm D}$. For simplicity, we require $a$, $b$ and $c$ to be
real and get the mass eigenvalues of $M^{}_{\rm R}$
\begin{eqnarray}
M^{}_{1} & = & \frac{a + b + 2c - \sqrt{ \left ( a -b \right )^{2}
+ 4 c^{2}}}{2 \left ( a b + b c + c a \right )} ~\Lambda^2_{\rm D}
\; , \nonumber \\
M^{}_{2} & = & \frac{a + b + 2c + \sqrt{ \left ( a -b \right )^{2}
+ 4 c^{2}}}{2 \left ( a b + b c + c a \right )} ~\Lambda^2_{\rm D}
\; .
\end{eqnarray}
Given $\Lambda^{}_{\rm D} \sim 174$ GeV (i.e., the scale of
electroweak symmetry breaking), $a \sim 0.022$ eV and $b \sim c
\sim 0.006$ eV as the typical inputs, the masses of two
right-handed Majorana neutrinos turn out to be $M^{}_{1} \sim 1
\times 10^{15}$ GeV and $M^{}_{2} \sim 3 \times 10^{15}$ GeV,
which are quite close to the energy scale of grand unified
theories $\Lambda^{}_{\rm GUT} \sim 10^{16}$ GeV. Note that the
textures of $M^{}_{\rm D}$ and $M^{}_{\rm R}$ taken in Eq. (18)
are by no means unique, but they may serve as a good example to
illustrate how the seesaw mechanism works to give rise to
$M^\prime_\nu$ or ${\cal L}^\prime_{\rm mass}$ in the MSM.

To summarize, we emphasize that the FL symmetry is a new kind of
flavor symmetry applicable to the building of neutrino mass
models. Imposing this symmetry on the effective Majorana neutrino
mass operator, we have shown that it can be broken in such a novel
way that the lightest neutrino remains massless but an
experimentally-favored bi-large neutrino mixing pattern is
achievable. This phenomenological scenario predicts a testable
relationship between the unknown neutrino mixing angle
$\theta^{}_{13}$ and the known angles $\theta^{}_{12}$ and
$\theta^{}_{23}$ in the CP-conserving case: $\sin\theta^{}_{13} =
\tan\theta^{}_{12} |(1- \tan\theta^{}_{23})/ (1+
\tan\theta^{}_{23})|$. Such a result is suggestive and interesting
because it directly correlates the deviation of $\theta^{}_{13}$
from zero with the deviation of $\theta^{}_{23}$ from $\pi/4$. We
have discussed a simple but instructive possibility of introducing
CP violation into the Majorana neutrino mass operator, in which
the soft breaking of $\mu$-$\tau$ permutation symmetry yields
$\delta = \pi/2$ (or $\delta = -\pi/2$) but keeps $\theta^{}_{23}
=\pi/4$. We have also discussed the possibility of incorporating
our scenario in the MSM.

In conclusion, the FL symmetry and its breaking mechanism may have
a wealth of implications in neutrino phenomenology. The physics
behind this new flavor symmetry remains unclear and deserves a
further study.

\vspace{0.4cm}

One of us (Z.Z.X.) likes to thank Z. Chang, X.D. Ji, M. Li, J.X.
Lu and S. Zhou for very useful discussions. This work was
supported in part by the National Natural Science Foundation of
China.

\newpage

\begin{figure}
\begin{center}
\vspace{0.5cm}
\psfig{file=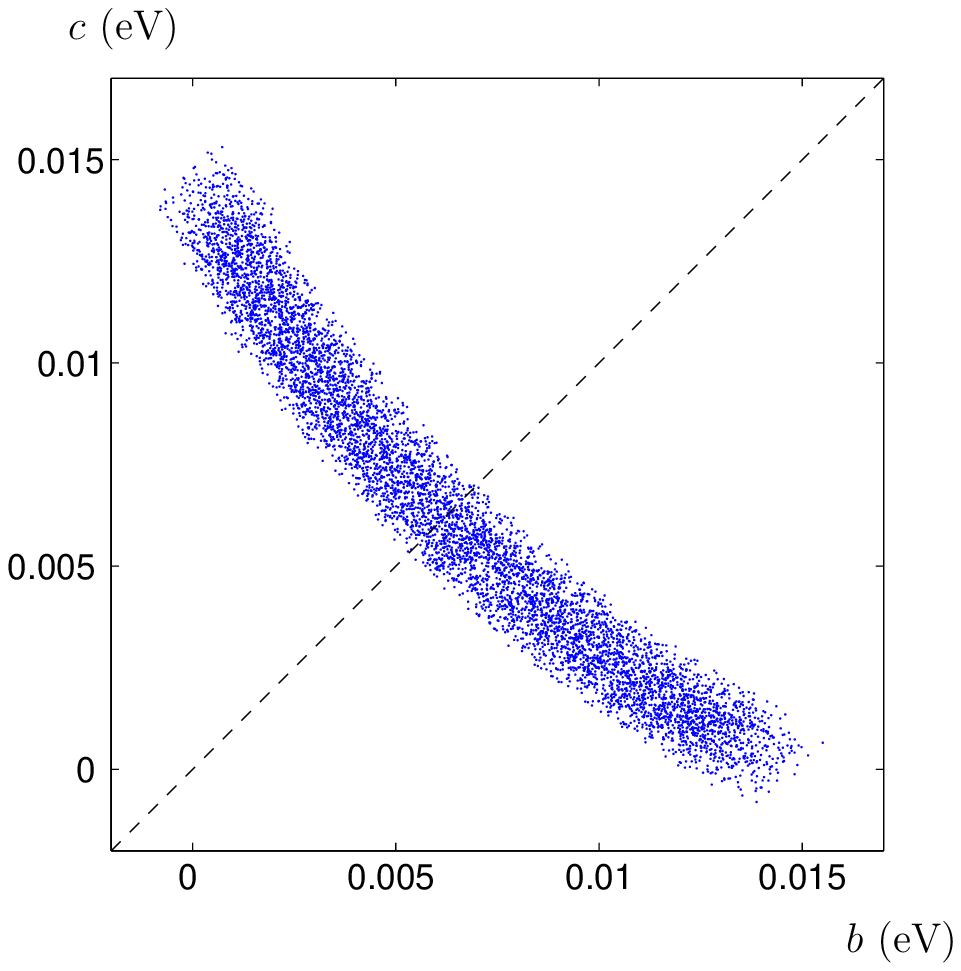, bbllx=2.2cm, bblly=6.0cm, bburx=12.2cm, bbury=16.0cm,%
width=7.5cm, height=7.5cm, angle=0,
clip=0}\vspace{1.5cm}\caption{The ranges of $b$ and $c$
constrained by current data through Eqs. (10), (11) and (13).}
\end{center}
\end{figure}

\begin{figure}
\begin{center}
\vspace{0.5cm}
\includegraphics[width=9cm,height=9cm]{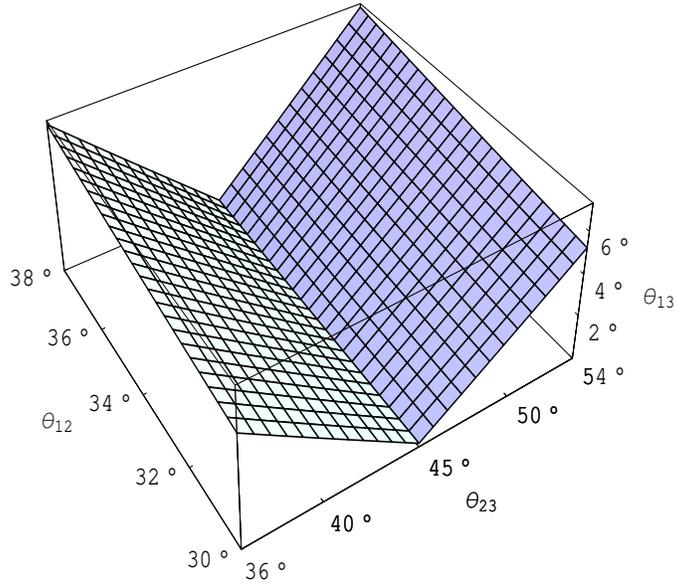}
\vspace{0cm} \caption{The numerical dependence of $\theta^{}_{13}$
on $\theta^{}_{12}$ and $\theta^{}_{23}$ as analytically predicted
by Eq. (14).}
\end{center}
\end{figure}

\end{document}